\documentclass{ws-ijmpcs}

\newcommand{\HQSS}{{\rm HQSS}}

\newcommand{\SU}{\mbox{SU}}
\newcommand{\U}{\mbox{U}}
\newcommand{\bc}{{\bar{c}}}
\newcommand{\bq}{{\bar{q}}}

\newcommand{\ben}{\begin{enumerate}}
\newcommand{\een}{\end{enumerate}}

\newcommand{\be}{\begin{equation}}
\newcommand{\ee}{\end{equation}}
\newcommand{\bea}{\begin{eqnarray}}
\newcommand{\eea}{\end{eqnarray}}
\newcommand{\ds}{\begin{displaystyle}}
\renewcommand{\ss}{\begin{scriptstyle}}

\newcommand{\ignore}[1]{}

\newcommand{\ba}{\begin{eqnarray}}
\newcommand{\ea}{\end{eqnarray}}

\begin{document}

\markboth{O. Romanets, C. Garc\'ia-Recio, J. Nieves, L. L. Salcedo, and L. Tol\'os}
{$N$ and $\Delta$ Hidden-Charm Resonances with Heavy-Quark Spin Symmetry}

%
\catchline{}{}{}{}{}
%

\title{$N$ AND $\Delta$ HIDDEN-CHARM RESONANCES WITH HEAVY-QUARK SPIN SYMMETRY
\footnote{Supported by Spanish Ministerio de Econom{\'\i}a y Competitividad
(FIS2011-28853-C02-02, FIS2011-24149, FPA2010-16963),
Junta de Andalucia (FQM-225), Generalitat Valenciana (PROMETEO/2009/0090) and EU HadronPhysics2 project (grant 227431). O. R. acknowledges support from the Rosalind Franklin Fellowship.
L. T. acknowledges support from RyC Program, and FP7-PEOPLE-2011-CIG (PCIG09-GA-2011-291679).}
}

\author{OLENA ROMANETS
}

\address{Theory~Group,~Kernfysisch~Versneller~Instituut,~University~of~Groningen, \\ Zernikelaan~25,
Groningen, 9747 ~AA, The~Netherlands
\\
o.romanets@rug.nl}

\author{CARMEN GARC\'IA-RECIO AND LORENZO LUIS SALCEDO}

\address{Departamento~de~F{\'\i}sica~At\'omica, Molecular~y~Nuclear, \\
 and Instituto
  Carlos I de F{\'\i}sica Te\'orica y Computacional, Universidad~de~Granada, \\
Granada,  E-18071, Spain} 

\author{JUAN NIEVES}
\address{
Instituto~de~F{\'\i}sica~Corpuscular~(centro~mixto~CSIC-UV),
  Institutos~de~Investigaci\'on~de~Paterna,\\
 Aptdo.~22085,~Valencia,~46071,~Spain} 

\author{LAURA TOL\'OS}
\address{ Institut~de~Ci\`encies~de~l'Espai~(IEEC/CSIC), 
  Campus~Universitat~Aut\`onoma~de~Barcelona, Facultat~de~Ci\`encies,
  Torre~C5,~Bellaterra,~E-08193,~Spain}  

\maketitle

\begin{history}
\received{3 August 2013}
\revised{18 September 2013}
\end{history}

\begin{abstract}
We study $N$ and $\Delta$ hidden-charm baryon resonances that are generated dynamically
from the $s$-wave interaction of pseudoscalar and vector mesons with $1/2^+$ and $3/2^+$ baryons.
We use a unitary coupled-channels model that fulfills heavy-quark spin symmetry and respects
spin-flavor symmetry in the light sector. 
We predict seven $N$-like and
five $\Delta$-like states with masses around 4~GeV, most of them as bound states. Some of these states
form heavy-quark spin multiplets, which are almost degenerate in mass. 

\end{abstract}

\ccode{PACS numbers: 14.20.Gk, 14.20.Pt, 11.10.St, 11.30.Ly}

\section{Introduction}	
The possible observation of new states with the charm degree of freedom 
has attracted a lot of attention over the past years in connection with past and 
on-going experiments such as CLEO, Belle, {\it BABAR}~\cite{experiments1}, 
and planned PANDA and CBM experiments at FAIR~\cite{experiments2}.
In this regard, it is important to understand the nature of possible new states,
e.g. whether baryon resonances can be interpreted as three-quark states or
molecular states.
The attention of the theoretical community has turned towards predicting 
features of possible hidden-charm states. 
Such baryon states were studied using zero-range vector exchange 
models~\cite{HofmannLutz}, hidden-gauge formalism~\cite{Oset}, and
constituent-quark model~\cite{Yuan:2012wz}.
Recently, hidden-charm baryon resonances were studied 
within the unitarized model in coupled channels, that uses the Weinberg-Tomozawa (WT) potential
extended to the $\SU(8)$ spin-flavor symmetry
and appropriately modified to respect heavy-quark spin symmetry (HQSS)~\cite{Paper}. 
Here we review some results of this last work.

\section{Theoretical Framework}
We start with the extension of the WT interaction to $\rm SU(8)$ 
spin-flavor symmetry. 
The corresponding Hamiltonian  for number of flavors $N_F$ and number of colors $3$
reads~\cite{GarciaRecio:2006wb}
\begin{equation}
{\mathcal H}_{\rm WT}^{\rm sf}(x)
= -\frac{{\rm i}}{4f^2} :[\Phi, \partial_0 \Phi]^A{}_B
{\cal B}^\dagger_{ACD} {\cal B}^{BCD}:
,
\quad
A,B,\ldots = 1,\ldots,2N_F
,
\label{eq:2.9}
\end{equation}
where $\Phi^A{}_B(x)$ is the meson field,  
which contains the fields of $0^-$ (pseudoscalar) and $1^-$ (vector)
mesons,
and
${\cal B}^{ABC}$ is a completely symmetric
tensor, which contains fields the lowest-lying baryons
with $J^P=\frac{1}{2}^+$ and $\frac{3}{2}^+$.
The Hamiltonian incorporates two distinct mechanisms:
%
the exchange part $H_{\rm ex}$, in which a quark is
transferred from the meson to the baryon, as another one is transferred from
baryon to meson; and the annihilation-creation $H_{\rm ac}$  mechanism, where
an antiquark in the meson annihilates with a similar quark
in the baryon, with subsequent creation of a quark and an antiquark.
The $H_{\rm ac}$ can violate 
HQSS
 when the annihilation or
creation of $q\bq$ pairs involve heavy quarks,
as
in the heavy-quark limit the number of charm quarks and the number of charm 
antiquarks are separately conserved (implying $\U_c(1)\times \U_\bc(1)$).
The HQSS group, which also includes a group of separate rotations of the $c$ quark and 
$\bar c$ antiquark, reads as
$\SU_c(2) \times \SU_\bc(2) \times \U_c(1)\times \U_\bc(1)$.
In the hidden-charm sectors, where in $H_{\rm ac}$ the annihilated and created antiquark 
is necessarily $\bar c$, we simply remove the offending annihilation-creation part of the Hamiltonian~\cite{Paper}.
 
As baryon resonances appear as poles of the scattering amplitude, we calculate 
the latter by solving the on-shell Bethe-Salpeter equation in coupled channels.
%
%
 The poles of the scattering amplitude on the first
Riemann sheet that appear on the real axis below threshold are 
bound states, and the ones in the second Riemann sheet below the
real axis and above threshold are identified with resonances
Often we
  refer to all poles generically as resonances, regardless of their concrete
  nature, since usually they can decay through other channels not included in
  the model space.
%
%

The WT potential with HQSS constraints possesses 
$\SU(6) \times \HQSS$ symmetry ($\SU(6)$ is the spin-flavor symmetry for the three light flavors). 
We break this symmetry adiabatically,
by implementing the physical hadron masses and meson decay constants,
 following the chain $\SU(6) \times \HQSS \to \SU(3) \times \HQSS \to \SU(2) \times \HQSS \to \SU(2)$,
where $\SU(3)$ is the flavor symmetry and the $\SU(2)$ is the isospin symmetry.
In this way we could follow the evolution of the poles while breaking the symmetry and
classify the found states under the corresponding group multiplets~\cite{Paper}.

\section{Dynamically-generated $N$ and $\Delta$ states}

\begin{figure*}[h]
\begin{center}
\includegraphics[scale=0.45]{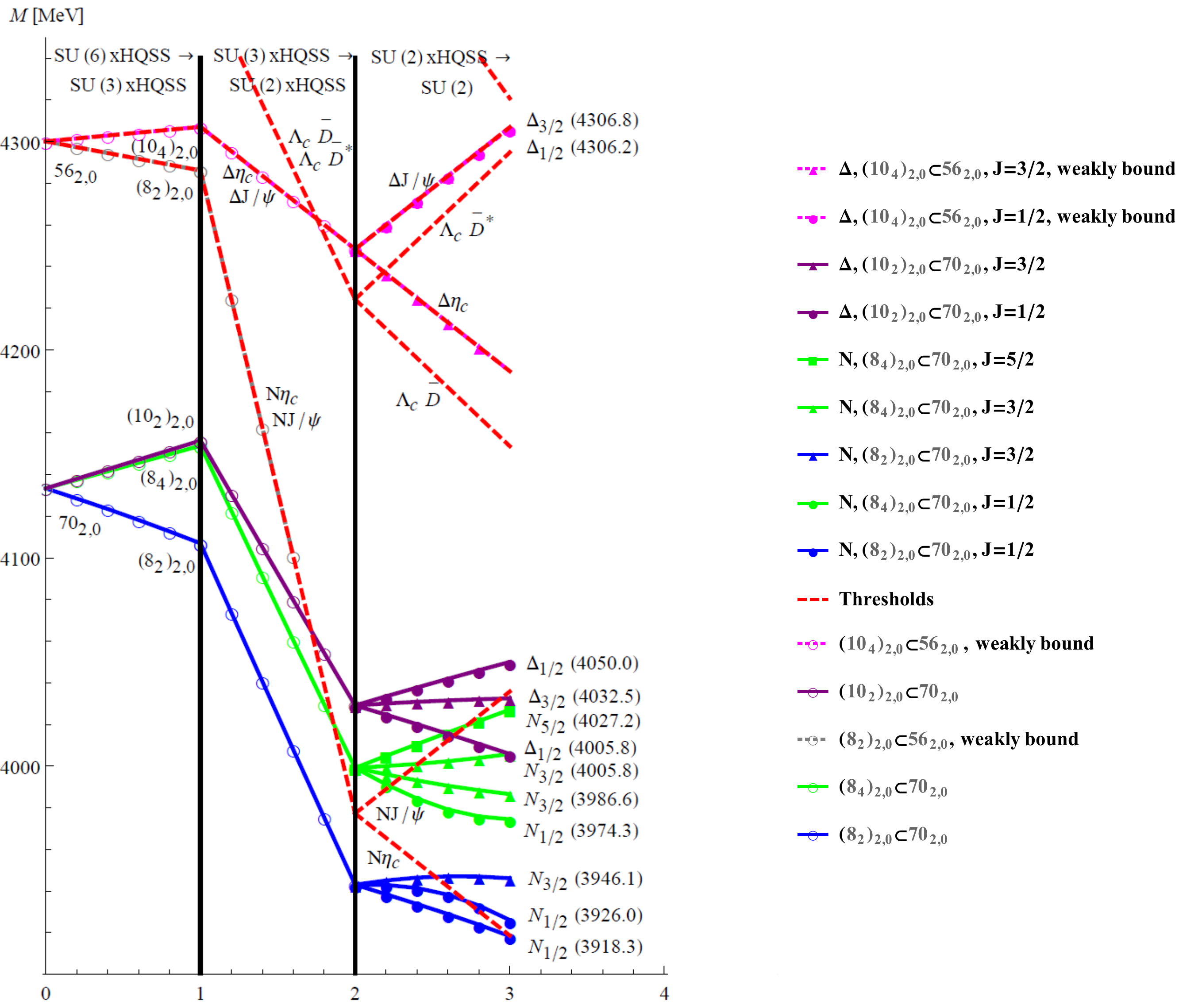}
\medskip
\vskip0.3cm 
\end{center}
\caption{Evolution of the poles as symmetries, starting from $\SU(6)\times \HQSS$,
  are sequentially broken to reach the isospin symmetric final crypto-exotic
  $N$ and $\Delta$ odd-parity resonances.  
The lower index of the final states
    stands for the spin $J$ of the corresponding resonance. The thresholds
    (red dashed lines) are marked together with the respective baryon-meson channel.
    The $\SU(6)\times\HQSS$ labels ${\bf 70_{2,0}}$ and ${\bf 56_{2,0}}$, and
    the $\SU(3 )\times\HQSS$ labels [${\bf (8_2)_{2,0} }$, ${\bf (8_4)_{2,0} }$,
    ${\bf (10_2)_{2,0} }$, ${\bf (10_4)_{2,0} }$] are also shown at the
    corresponding symmetric points. } 
\label{fig_evolution}
\end{figure*}

There are two attractive representations in the hidden-charm sector with $C=0$,
$\bf{56_{2,0}}$ and $\bf{70_{2,0}}$, where the first index stands for $2 J_c+1$, with
$J_c$ being the spin of the charm quark, and the second index is the total charm. 
The evolution of the poles that correspond to the $N$ and $\Delta$ resonances is shown
Fig.~\ref{fig_evolution}. The masses of the final states and their spins $J$ can also be found on the 
figure.
We find three $N_{1/2}$ (the lower index indicates $J$), three $N_{3/2}$, 
and one $N_{5/2}$, with masses between 3918 and 4027 MeV. Some 
of the states are degenerated when the HQSS is unbroken, thus forming
the HQSS multiplets, as can be seen from the Fig.~\ref{fig_evolution}.
Almost all found $N$ resonances are bound states, excluding $N_{1/2}(3926)$
which has a small width of 0.1~MeV, and $N_{1/2}(3974)$, with a width of
2.8~MeV.

The $N$ resonances studied in the zero-range vector exchange model~\cite{HofmannLutz}
are about 500 MeV lighter than those found in our model.
%
The hidden-gauge
formalism predicts these masses to be about 400 MeV larger~\cite{Oset},
however this difference comes mostly from using a different renormalization
prescription~\cite{XiaoNievesOset}.
Our results are close to those predicted by the chiral interaction, studied in the constituent 
quark model of Ref.~\refcite{Yuan:2012wz}, whereas the instanton-induced interaction
and color-magnetic interaction produce higher masses for the resonances~\cite{Yuan:2012wz}.

Further, we find three $\Delta_{1/2}$ and two $\Delta_{3/2}$ 
bound states. Two of them, $\Delta_{1/2}(4306)$ and $\Delta_{3/2}(4307)$,
which stem from the $\bf 56_{2,0}$ representation, appear as cusps in the scattering amplitude.

\section{Summary}
%
We use a suitable broken $\SU(8)$ spin-flavor extended WT potential for studying
the hidden-charm $N$ and $\Delta$ resonances. Heavy-quark symmetry is enforced by
removing heavy quark-antiquark annihilation-creation mechanisms that would
violate this symmetry.
Our model
predicts the existence of seven $N$ states, and five $\Delta$ states,
with masses around 3.9-4.3~GeV. These results can be compared with the predictions of other models,
and can be tested in the future PANDA and CBM experiments at FAIR facility.

\section{References}



\begin{thebibliography}{0}    

\bibitem{experiments1}
www.lepp.cornell.edu/Research/EPP/CLEO,{\enskip}belle.kek.jp,
www-public.slac.stanford.edu/babar.

\bibitem{experiments2}
http://www.fair-center.eu/.

\bibitem{HofmannLutz} 
  J.~Hofmann and M.~F.~M.~Lutz,
  Nucl.\ Phys.\ A {\bf 763}, 90 (2005); 
  Nucl.\ Phys.\  A {\bf 776}, 17 (2006).

\bibitem{Oset}
  J.~-J.~Wu, R.{\it et al.},
  Phys.\ Rev.\ C {\bf 84}, 015202 (2011);
  J.~-J.~Wu {\it et al.},
  Phys.\ Rev.\ C {\bf 85}, 044002 (2012).

\bibitem{Yuan:2012wz}
  S.~G.~Yuan
{\it et al.}
  Eur.\ Phys.\ J.\ A {\bf 48},61 (2012).


\bibitem{Paper}
 C.~Garcia-Recio, J.~Nieves, O.~Romanets, L.~L.~Salcedo, and L.~Tolos,
  Phys.\ Rev.\ D {\bf 87}, 074034 (2013).

\bibitem{GarciaRecio:2006wb} 
  C.~Garcia-Recio, J.~Nieves and L.~L.~Salcedo,
  Phys.\ Rev.\ D {\bf 74}, 036004 (2006).

\bibitem{XiaoNievesOset}
C.~W.~Xiao, 
 J.~Nieves, E.~Oset,
 arXiv:1304.5368 [hep-ph].


\end{thebibliography}
\end{document}